\documentclass[twocolumn,showpacs,amssymb,aps,prl,floatfix]{revtex4-1}
                                              
\usepackage{multirow}
\usepackage{booktabs}
\usepackage{latexsym} 	
\usepackage{amsmath}  	
\usepackage{amsbsy}
\usepackage{epsfig}     
\usepackage{graphicx}
\usepackage{latexsym}
\usepackage{euscript}
\usepackage{amsfonts}
\usepackage{mathtools}
\usepackage{braket}
\usepackage{simplewick}
\usepackage{amsmath}
\usepackage{amssymb}
\usepackage{dsfont}
\usepackage{slashed}
\graphicspath{{fig/}}

\newcommand\tr{\,{\rm Tr}\,}
\newcommand{\re}{\,{\rm Re }\,}

\newcommand{\om}{\omega}

\newcommand{\vecnul}{{\mathbf 0}}

\newcommand{\pv}{{\mathbf p}}

\newcommand{\xv}{{\mathbf x}}

\newcommand{\be}{\begin{equation}}
\newcommand{\ee}{\end{equation}}
\newcommand{\bea}{\begin{eqnarray}}
\newcommand{\eea}{\end{eqnarray}}
\newcommand{\bean}{\begin{eqnarray*}}
\newcommand{\eean}{\end{eqnarray*}}
\newcommand{\nn}{\nonumber}

\begin{document}

\title{Electrical conductivity of the quark-gluon plasma across the deconfinement transition 
}

\author{Alessandro Amato$^{1,2}$}
\author{Gert Aarts$^1$}
\author{Chris Allton$^1$}
\author{Pietro Giudice$^{1,3}$}
\author{Simon Hands$^1$}
\author{Jon-Ivar Skullerud$^4$}

\affiliation{$^1$Department of Physics, College of Science, Swansea University,
Swansea SA2 8PP, United Kingdom}
\affiliation{$^2$Institute for Theoretical Physics, Universit\"at Regensburg, D-93040 Regensburg, Germany}
\affiliation{$^3$Institut f\"ur Theoretische Physik, Universit\"at M\"unster,  D-48149, M\"unster, Germany}
\affiliation{$^4$Department of Mathematical Physics, National University of
Ireland Maynooth, Maynooth, Co Kildare, Ireland}

\date{July 25, 2013}

\begin{abstract}
A lattice calculation is presented for the electrical conductivity $\sigma$ of the QCD
plasma with $2+1$ dynamical flavours at nonzero temperature.
We employ the conserved lattice current on anisotropic lattices using a tadpole-improved clover action and study the behaviour of the conductivity over a wide range of temperatures, both below and above the deconfining transition.
The conductivity is extracted from a spectral-function analysis using the Maximal Entropy Method, and a discussion of its systematics is provided. 
We find an increase of $\sigma/T$ across the transition.
\end{abstract}


\maketitle


{\em Introduction} --
 Transport coefficients, such as the shear and bulk viscosity, the electrical conductivity and heavy-quark diffusion constants, can be seen as  parameters in the low-energy effective
theories that describe real-time evolution in the quark-gluon plasma (QGP) on the longest length and time scales,
 encoding the dynamics of the underlying quantum field theory, QCD.
Knowledge of transport coefficients is especially relevant for understanding heavy-ion collision experiments, at the Relativistic Heavy-Ion Collider at BNL and the Large Hadron Collider at CERN, for which viscous hydrodynamics is routinely used as a tool to analyse the collisions \cite{Heinz:2013th,Gale:2013da}.

In the strongly coupled quark-gluon plasma, transport coefficients are not easily calculable. 
Although the application of holography in strongly coupled theories has provided an important stimulus \cite{Kovtun:2004de}, a nonperturbative computation directly in QCD is highly desirable. 
According to the Kubo formulas, see e.g.\ Ref.\ \cite{Kadanoff1963419}, transport coefficients can be
extracted from the low-energy behaviour of appropriate current-current spectral functions.
When using lattice QCD, which is formulated in euclidean space-time, the main challenge is the construction of spectral functions from euclidean correlators, i.e.\ the analytical continuation from imaginary to real time. Here the understanding has steadily increased and some
lattice results for transport coefficients (mostly in SU(3) gauge theory, i.e.\ without dynamical quarks) are now available \cite{Gupta,Aarts:2007wj,Meyer:2007ic,Meyer:2007dy,Ding:2010ga,brandt}.
We discuss this further below and refer to 
Refs.\  \cite{aarts,Schafer:2009dj,meyer} for reviews on transport which include a discussion of lattice QCD aspects. 

In this paper we present lattice results for the electrical conductivity $\sigma$ of the QCD plasma. On the phenomenological side the conductivity may play an important role in the evolution of electromagnetic fields during a heavy-ion collision  \cite{Tuchin:2013ie,McLerran:2013hla} and it has recently also been suggested that experimental information on conductivity  can be extracted from flow parameters in heavy-ion collisions \cite{Hirono:2012rt}.
There are a number of lattice QCD computations of the conductivity, using a plasma without dynamical quarks (quenched, $N_f=0$) \cite{Gupta,Aarts:2007wj,Ding:2010ga}. A recent two-flavour study at a single temperature $T$ in the QGP is available as well \cite{brandt}. Here we improve on these results in various significant ways. First of all, our simulations are carried out in a plasma with $N_f=2+1$ quark flavours. Secondly, we consider a wide range of temperatures below and above the deconfinement transition. This allows us for the first time to observe a rise of the conductivity as the temperature is increased \cite{Francis:2011bt}.
Third, we use the exactly conserved current on the lattice, whereas in all previous studies a local lattice operator, which requires renormalisation, was used. And finally, we employ anisotropic lattices with a substantially smaller lattice spacing in the time direction, allowing for more data points to be used in the analysis.
Our main result is the observation of an increase of $\sigma/T$, as the plasma is heated from the confined to the deconfined phase.


{\em Electrical conductivity} --
 The electromagnetic (em) current is given by 
$j^{\rm em}_\mu(x) = e\sum_f q_f j^f_\mu(x)$,
 where the sum is over the flavours, $e$ is the elementary charge, $q_f$ is the fractional charge of the quark ($2/3$ or $-1/3$) and $j_\mu^f$ is the vector current for each flavour.
The euclidean correlator $G^{\rm em}$ built up from $ j^{\rm em}_\mu$ is related
to the corresponding spectral function $\rho$ via the integral relation  \cite{aarts,Schafer:2009dj,meyer} 
\begin{align}
\nn
 G^{\rm em}_{\mu\nu}(\tau, \pv) = & \int d^3x\, e^{i \pv \cdot \xv  }
 \braket{\,j^{\rm em}_\mu(\tau,\xv) j^{{\rm em}}_\nu(0,\vecnul)^\dagger\,} \\
 = & \int_0^\infty \frac{d\omega}{2\pi}\, K(\tau,\omega)\,\rho^{\rm em}_{\mu\nu}(\omega,\pv),
  \label{spectr}
\end{align} 
where the temperature-dependent kernel is given by
\be
\label{kernel}
 K(\tau,\omega) = \frac{\cosh[\omega(\tau-1/2T)]}{\sinh[\omega/2T]}.
\ee
In the rest of the paper we consider correlators at zero momentum and hence we drop the $\pv$ dependence.
The electrical conductivity $\sigma$ can finally be determined from the slope of 
the spectral function at $\om=0$ as \cite{Kadanoff1963419}
\be
 \label{cond}
 \frac{\sigma}{T} = \frac{1}{6T}  \lim_{\omega \rightarrow 0}\frac{\rho^{\rm em}(\omega)}{\omega}, 
 \quad\quad
 \rho^{\rm em}(\om) = \sum_{i=1}^3 \rho^{\rm em}_{ii}(\om).
\ee


{\em Lattice details} --
 We use an anisotropic lattice of size $N_s^3\times N_\tau$
with 2+1 flavours of clover fermions \cite{clover,anis1,anis2,chen}. 
Using a finer lattice spacing $a_\tau$ for the time direction provides a better temporal 
resolution of the correlation functions without increasing the
computational cost significantly.
However this choice introduces new bare parameters in the action, which have to be
tuned carefully. This has been achieved in Refs.\  \cite{anis2,anis1}, to which we refer for further details. The gauge action is   Symanzik-improved with tree-level tadpole-improved coefficients. 
The Dirac operator reads \cite{fn1}
\begin{eqnarray}
 \label{clover}
 D[U] &=& 
  \hat{m}_0 +  \gamma_4 \hat{W}_4 +\frac{1}{\gamma_f}  \sum_i \gamma_i \hat{W}_i \nonumber
\\
 && - \frac{c_t}{2} \sum_{i}\sigma_{4i}\hat{F}_{4i} 
 -  \frac{c_s}{2\gamma_g}\sum_{i<j} \sigma_{ij} \hat{F}_{ij}.
\end{eqnarray}
The first three terms indicate the mass term and the usual Wilson operator $\hat W_\mu$, with $\gamma_\mu$ the Dirac matrices, while the last two  are the clover operators, 
with $\sigma_{\mu\nu}  =  \frac{i}{2}[\gamma_{\mu}, \gamma_{\nu}]$ and
$\hat{F}_{\mu\nu}$ the lattice version of the field strength tensor. 
The parameters $\gamma_g$  and $\gamma_f$ are the bare gauge and fermion anisotropies, which have to be tuned.
  Following Ref.\ \cite{anis2} we use a renormalised anisotropy $ \xi \equiv a_s / a_\tau = 3.5$, which results from $\gamma_g=4.3$ and $\gamma_f=3.4$. 
The parameters in front of the timelike and spacelike clover operators, $c_t=0.9027$ and
$c_s=1.5893$, have been chosen according to tree-level conditions \cite{chen}.
The gauge links $U_\mu$ are represented by three-dimensional stout-smeared links
\cite{smear}, with smearing weight $\rho=0.14$ and $n_\rho=2$ iterations.
The light and strange quark mass parameters are chosen  \cite{anis1} to reproduce the physical strange quark mass and a light quark mass with $M_\pi/M_\rho = 0.446(3)$, i.e. $\hat{m}_0^{\rm light} = -0.0840$ and  $\hat{m}_0^{\rm strange} = - 0.0743$.

In Refs.\ \cite{anis2,anis1} only zero-temperature lattices were considered, with $a_\tau N_\tau\gg a_sN_s$ and $N_s=12,16,24$. We have generated a number of finite-temperature ensembles, using spatial lattice extents of $N_s=24$ and 32, and $N_\tau$ ranging from 48 to 16. Some details are given in Table \ref{tab:lattice}. As always, the temperature is given by $T=1/(a_\tau N_\tau)$. The  critical temperature is estimated from the renormalised Polyakov loop inflection point \cite{inprep1}. Note that we have four temperatures both below and above $T_c$.

In order to compute the electromagnetic current correlator, we use the exactly conserved vector current on the lattice, whose components at lattice site $x$ read
\begin{align}
 V_{\mu}^\text{C}(x) = \kappa_\mu \bigg[ & \bar\psi(x+\hat\mu)(1+\gamma_\mu) \, U_\mu^\dagger(x) \, \psi(x) 
  \nonumber \\ 
 & - \bar\psi(x) (1-\gamma_\mu) \, U_\mu(x) \, \psi(x+\hat\mu)\bigg], 
\end{align} 
where $\kappa_4=1/2$, $\kappa_i=1/(2\gamma_f)$.
To compute the current-current correlator as a
function of the euclidean time separation $\tau$ in the zero-momentum limit, we use Wick's contraction and neglect disconnected diagrams, as has been the case in all previous studies and is well motivated \cite{Ding:2010ga}: in particular their contribution is identically zero in the $N_f=3$ case.
We then find the following two contributions to the correlator,
\begin{align}
& \braket{ V_\mu^\text{C}(x)\,V_\nu^\text{C}(y)^\dagger }= 2\kappa_\mu\kappa_\nu \re\tr\big[ 
\nn  \\
  -
&S(y,x+\hat\mu) \, U_\mu^\dagger(x) \, \Gamma^+_\mu \, S(x,y+\hat\nu) \, U_\nu^\dagger(y) \, \widetilde\Gamma^-_\nu
\nn \\
 +
&S(y+\hat\nu,x+\hat\mu) \, U_\mu^\dagger(x) \, \Gamma^+_\mu \, S(x,y) \, U_\nu(y) \, \widetilde\Gamma^+_\nu \; \big],
\end{align} 
 where $x,y$ are two lattice points,
$S(x,y)=\braket{\psi(x)\bar\psi(y)}$ is the fermion propagator
and $\Gamma^\pm_\mu=1\pm\gamma_\mu$,
$\widetilde\Gamma^\pm_\mu=1\pm\widetilde\gamma_\mu$ with
$\widetilde{\gamma}_\mu=\gamma_4 \gamma_\mu \gamma_4$.
Since we are using a Symanzik-improved action,  the current should in principle be improved as well, by adding a total divergence of the form $a_\nu \partial_\nu \bar \psi(x) \,\sigma_{\mu\nu}\,\psi(x)$, which we have not done (for massless quarks this contribution is suppressed).
To convert from lattice to continuum units on an anisotropic lattice, we note that the spatial current density $V_i^C(x)$  is given in units of $a_s^2 a_\tau$. It then follows that the correlator, projected to zero momentum, and its spectral function,  are given in units of $a_sa_\tau^2$ and $a_sa_\tau$ respectively.
Finally, in order to be able to compare with results obtained previously, we consider in this paper only the contribution from the two light flavours to the electromagnetic current. We can then factor out the fractional charge assignments of the quarks, via $C_{\rm em} = e^2\sum_f q_f^2= 5/9e^2$, and write
$\rho^{\rm em}(\om) = C_{\rm em}\rho(\om)$.
 The inclusion of the strange quark is in progress \cite{inprep2}.

\begin{table}[t]
 \begin{center}
   \begin{tabular}{ccccccc}
    \hline\hline
    \ $N_s$\ &\ $N_\tau$\  &\ $T$\ [MeV] &\ $T/T_c$\  &\ $N_{\texttt{CFG}}$\ &\ $N_{\texttt{SRC}}$\  \\
\hline\hline
   32 &   16 & 352 & 1.90  & 1059 &4  \\ 
   24 &   20 & 281 & 1.52 & 1001 &4 \\ 
   32 &   24 & 235 & 1.27 & 500   &4 \\ 
   32 &   28 & 201 & 1.09 & 502   &4  \\ 
   32 &   32 & 176 & 0.95 & 501   &4  \\ 
   24 &   36 &156  & 0.84 & 501   &4  \\
   24 &   40 &141  & 0.76 & 523   &4  \\
   32 &   48 &117  & 0.63 & 601   &1 \\ \hline\hline
    \end{tabular}
\end{center}
\caption{Lattice setup: the lattice size is $N_s^3\times N_\tau$ and the lattice spacing is $a_s=0.1227(8)$ fm and $a_\tau=0.03506(23)$ fm, corresponding to $a_\tau^{-1}=5.63(4)$ GeV \cite{anis1,anis2}. The renormalised anisotropy is $\xi\equiv a_s/a_\tau = 3.5$.
 $N_{\texttt{CFG}}$ is the number of configurations available for each volume and $N_{\texttt{SRC}}$ is the number of sources used in the analysis. 
 }
\label{tab:lattice}
\end{table}


\begin{figure}[t]
 \includegraphics[width=0.4\paperwidth]{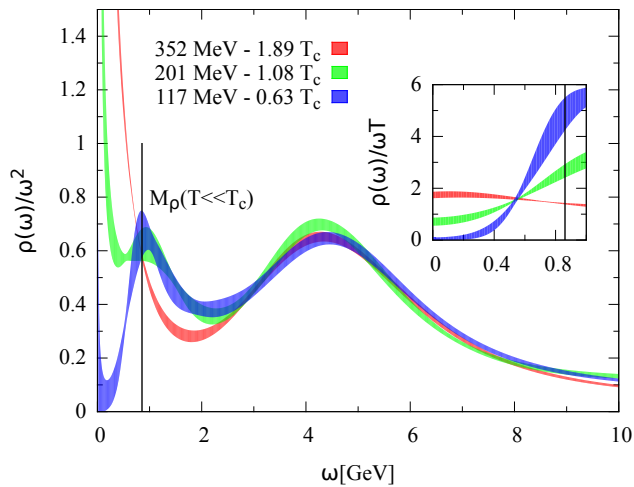}
\caption{
 Spectral functions $\rho(\om)/\omega^2$ and $\rho(\om)/\omega T$ (inset) at three  
temperatures. The vertical line indicates $M_\rho$ at $T=0$ \cite{anis1}, and the thickness of the lines represents the statistical jackknife error from MEM. The rise of the intercept at $\om=0$ in the inset indicates a temperature-dependent  conductivity. 
 }
 \label{fig:examplemem}
\end{figure}

{\em Spectral function} --
 The inversion of the integral equation (\ref{spectr}) relating $G(\tau)$ and $\rho(\om)$ is an ill-posed problem, since the correlator is known numerically at a finite number of time slices only, whereas the spectral function is, in principle, a continuous function of $\om$. 
To resolve this problem, one possibility is to use an Ansatz for $\rho(\om)$ with a small number of fitting parameters and a constrained fitting procedure required for stabilisation \cite{Ding:2010ga,brandt}. 
Another route is to use the Maximum Entropy Method (MEM) \cite{Asakawa:2000tr}, which has its basis in Bayesian analysis \cite{bayes} and aims to construct the most probable spectral function, given the data and prior information, encoded in a default model, without requiring any assumptions of its functional form.
 Here we use Bryan's algorithm \cite{bryan}, which expands $\rho(\om)$ in terms of basis functions determined by a singular-value decomposition of the kernel $K(\tau,\om)$ \cite{Asakawa:2000tr,Aarts:2007wj}.
At $T>0$, a straightforward implementation of this leads to instabilities since the kernel diverges as $\om\to 0$,  $K\sim1/\om$. To cure this we construct instead  $\rho(\om)/\om$, using the kernel $\om K(\tau,\om)$ \cite{Aarts:2007wj}. Prior information is then introduced via the default model $m(\om)$ as
\be
\label{eqrho}
\frac{\rho(\om)}{\om} = m(\om) \exp\sum_k c_k u_k(\om),
\ee
 where $u_k(\om)$ are the basis functions mentioned above and the coefficients $c_k$ are to be determined. 
We employ a default model with a minimal amount of features, i.e.,
\be
\label{default}
m(\omega)=m_0 (b + a_\tau\omega).
\ee
Here $m_0$ is an overall normalisation, determined by a simple $\chi^2$ fit to the correlator. The parameter $b$ is essential at small energies, since it permits the presence of a nonzero value of $\rho(\om)/\om$ at $\om=0$ and hence a nonzero conductivity. Varying $b$ provides a crucial test to verify the robustness of our results, as we will see below. 
Finally, the term linear in $\om$ is motivated by the expected large $\om$ behaviour in the continuum.

\begin{figure}[t]
 \includegraphics[width=0.4\paperwidth]{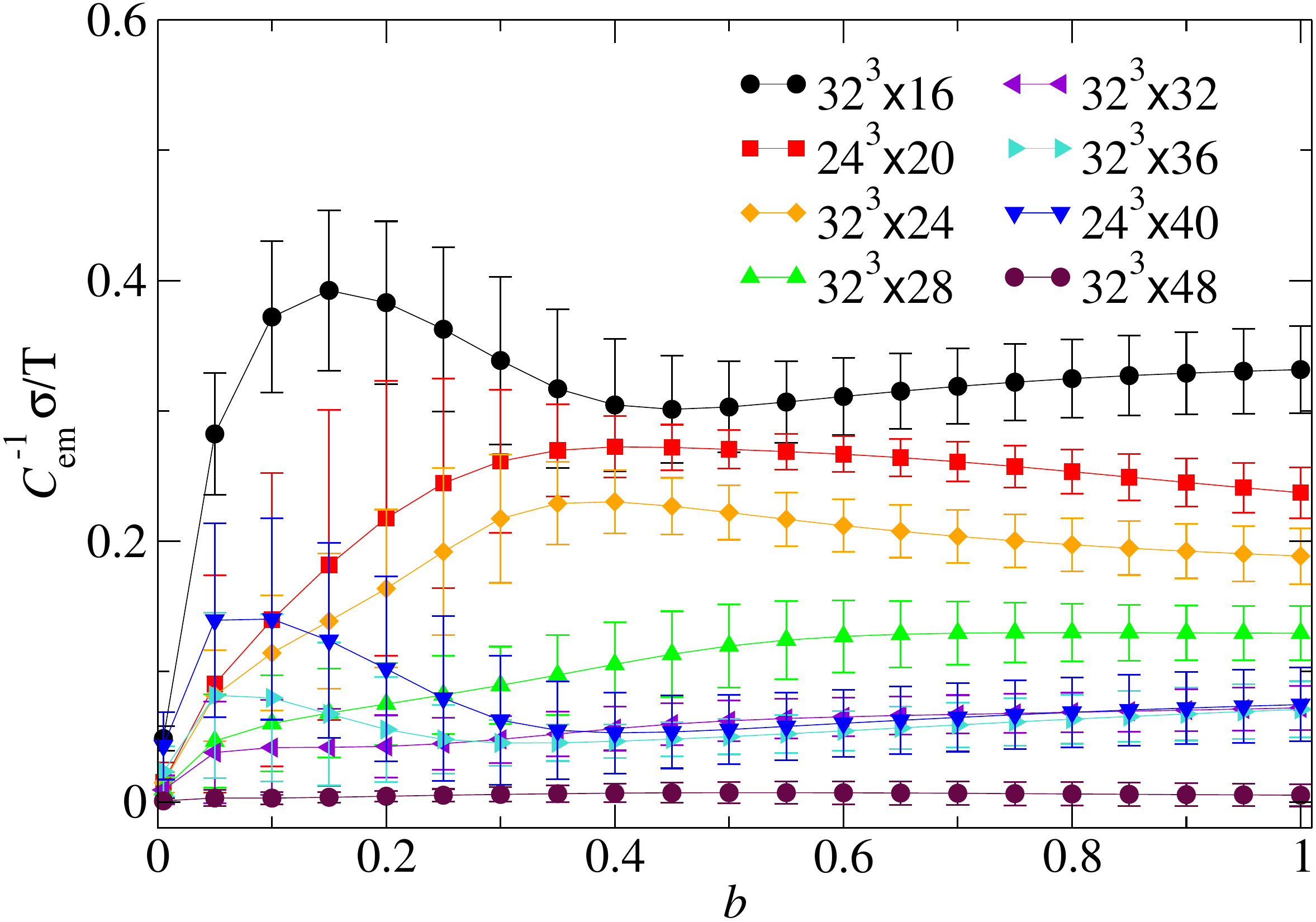}
 \caption{ 
 Dependence of the conductivity $C_{\rm em}^{-1}\sigma/T$ on the parameter $b$ in the default model (\ref{default}). The time range included in the MEM analysis is $\tau/a_\tau=4, \ldots, N_\tau/2$ and the $\omega$ range is $0<a_\tau \om<3$.
The result is robust against variations of $b$, provided it is not too small. 
 }
 \label{fig:bdep}
 \vspace*{-0.2cm}
\end{figure}


{\em Results} --
 We now discuss our results. 
 In Fig.\ \ref{fig:examplemem} we present spectral functions obtained with MEM for three temperatures.
The main figure shows $\rho(\om)/\omega^2$. At the lowest temperature we observe a peak in the spectral function corresponding to the $\rho$ particle. Note that the vertical line denotes $M_\rho$ at $T=0$ \cite{anis1}. As the temperature is increased, this peak is reduced and eventually disappears, which is interpreted as ``melting''. The structures at $\om\sim 4-6$ GeV are presumably lattice artefacts due to the finite size of the Brillouin zone and are not physical \cite{martinez}. In the inset, we show $\rho(\om)/\om T$, in order to highlight the presence of an intercept at $\om=0$. It can be seen clearly that as the temperature is increased, a nonzero intercept emerges, indicating the presence of a temperature-dependent conductivity.
Underlying this analysis is the assumption that the transport peak is not extremely narrow; if it is the inversion will not determine the intercept reliably \cite{Aarts:2002cc}.

In order to study the robustness of the MEM results, we have carried out a number of tests.  By varying the time interval included in the MEM analysis, we found that the results are stable when  $\tau_{\rm min} \leq \tau \leq a_\tau N_\tau/2$,
with  $\tau_{\rm min}/a_\tau \gtrsim 3$; the results shown here are obtained with $\tau_{\rm min}/a_\tau=4$.
Similarly, we varied the $\om$ range with $0\leq \om\leq \om_{\rm max}$ and found stability provided $a_\tau\om_{\rm max}\sim 3 - 5$; here we use $a_\tau\om_{\rm max}=3$.
An important test concerns the parameter $b$ in the default model, since this parameter is directly related to the intercept of $\rho(\om)/\om$ and hence $\sigma/T$, see Eqs.\ (\ref{cond}, \ref{eqrho}, \ref{default}). The $b$ dependence is shown in Fig.~\ref{fig:bdep}. We observe clear plateaus, provided that $b$ is not too small. In the latter case, the conductivity is unnaturally pushed to zero, due to a bias in the default model, which should be avoided. We also note a larger sensitivity to $b$ at the highest temperature, which reflects that in this case only a small number of time slices is available for the analysis.

\begin{figure}[t]
  \includegraphics[width=0.4\paperwidth]{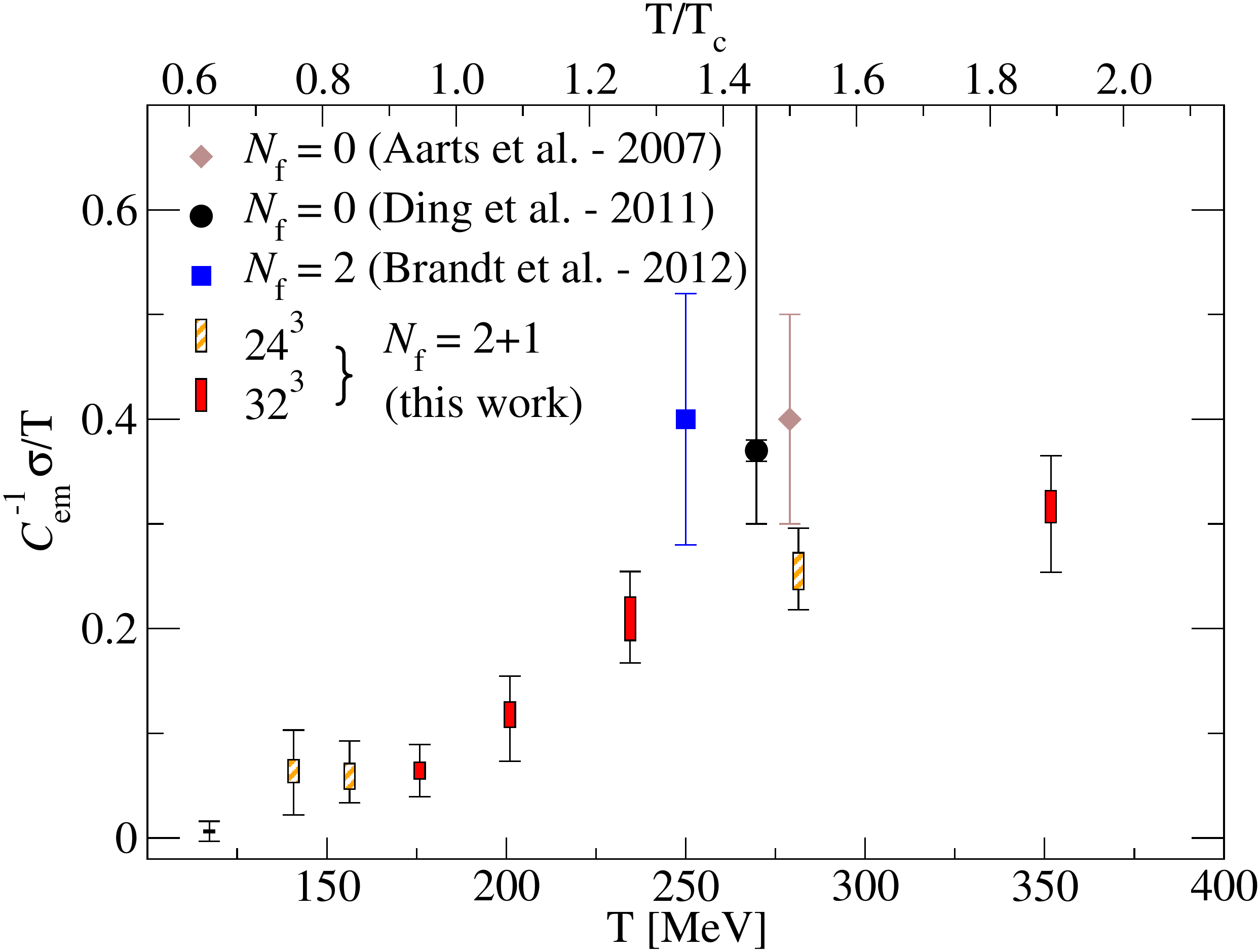}
 \caption{ Temperature dependence of $C_{\rm em}^{-1}\sigma/T$, where $C_{\rm em}=5/9e^2$ for two light flavours. The vertical size of the rectangles 
  reflects the systematic uncertainty due to changes in the default model,  by varying $0.4<b<1$. 
 The error bars indicate the statistical jackknife error, combining all $b$ values between 0.4 and 1.
  Previously obtained results  \cite{Aarts:2007wj,Ding:2010ga,brandt} are indicated as well: the $N_f=0$ results are inserted matching the values of $T/T_c$. Note that the black circle has two error bars \cite{Ding:2010ga}.
 }
 \label{fig:final}
  \vspace*{-0.1cm}
\end{figure}

Our results for the conductivity are shown in Fig.~\ref{fig:final} where $C^{-1}_{\rm em} \sigma / T$ is plotted against the temperature. 
We observe an increase of $\sigma/T$ as the transition to the deconfined phase is made, with the rise starting already below $T_c$. We note that since the transition is a crossover, a smooth transition may be expected. 
It is not excluded that far below $T_c$, the conductivity is much larger due to the transport of charged hadrons, which may, however, lead to a narrow transport peak, whose details cannot be resolved in the euclidean correlator  \cite{Aarts:2002cc}.
 Some previously obtained results are shown as well. We observe that our $N_f=2+1$ findings are comparable with those well inside the QGP phase.  
 Not shown are the much larger value, $\sigma/T\sim 7$, found in Ref.\ \cite{Gupta} above $T_c$, and the much smaller lower bound found in Ref.\ \cite{Burnier:2012ts} from a re-analysis of the data of Ref.\ \cite{Ding:2010ga}.


{\em Conclusion} --
 We have presented the first lattice QCD analysis of the electrical conductivity in the QCD plasma across the deconfinement  transition. While inside the QGP our results are comparable with previously obtained results, we have for the first time observed an increase of $\sigma/T$, starting already in the confined phase. 
  It would be interesting to explain this behaviour in effective QCD models or semi-analytically, both below and above $T_c$, see e.g.\  Refs.\ \cite{FernandezFraile:2005ka,Nam:2012sg,Cassing:2013iz,qin}. 
 In the near future, we plan to include  the contribution from the strange quark to the current.
 Finally, we note that this calculation only offers the QCD contribution to the conductivity and not the contribution from weakly interacting leptons.
  

\begin{acknowledgments}
 We thank Harvey Meyer for discussion.
 This work is undertaken as part of the UKQCD collaboration and the STFC funded DiRAC Facility. 
  We acknowledge the PRACE Grants 2011040469 and Pra05\_1129, European Union Grant Agreement No.\ 238353 (ITN STRONGnet), the Irish Centre for High-End
Computing, STFC, the Wolfson Foundation and the Royal Society for support.
  This work used the Chroma software suite  \cite{Edwards:2004sx}. 

\end{acknowledgments}



\begin{thebibliography}{10}
\bibitem{fn1}
Note that $\tilde u_t=1$ and carets indicate lattice units \cite{anis2,anis1}.
\end{thebibliography}
\end{document}